\documentclass[twocolumn,journal]{IEEEtran}
\usepackage[T1]{fontenc}
\usepackage[latin9]{inputenc}
\usepackage{color}
\usepackage{amsmath}
\usepackage{amsthm}
\usepackage{amssymb}
\usepackage{graphicx}
\usepackage[unicode=true,
 bookmarks=true,bookmarksnumbered=true,bookmarksopen=true,bookmarksopenlevel=1,
 breaklinks=false,pdfborder={0 0 0},pdfborderstyle={},backref=false,colorlinks=false]
 {hyperref}
\hypersetup{pdftitle={Your Title},
 pdfauthor={Your Name},
 pdfborderstyle={},pdfpagelayout=OneColumn,pdfnewwindow=true,pdfstartview=XYZ,plainpages=false}

\makeatletter

\providecommand{\tabularnewline}{\\}

\let\oldforeign@language\foreign@language
\DeclareRobustCommand{\foreign@language}[1]{%
  \lowercase{\oldforeign@language{#1}}}
\theoremstyle{plain}
\newtheorem{thm}{\protect\theoremname}
\theoremstyle{plain}
\newtheorem{lem}[thm]{\protect\lemmaname}

\usepackage[noadjust]{cite}

\providecommand{\lemmaname}{Lemma}
\providecommand{\theoremname}{Theorem}

\providecommand{\lemmaname}{Lemma}
\providecommand{\theoremname}{Theorem}

\makeatother

\providecommand{\lemmaname}{Lemma}
\providecommand{\theoremname}{Theorem}

\begin{document}
\title{Generalized BER of MCIK-OFDM with Imperfect CSI: Selection combining
GD versus ML receivers}
\author{Vu-Duc Ngo,~Thien Van Luong, Nguyen Cong Luong, Minh-Tuan Le, Thi
Thanh Huyen Le, and Xuan-Nam Tran\thanks{V.-D. Ngo is with the School of Electronics and Electrical Engineering,
Hanoi University of Science and Technology, Hanoi 11657, Vietnam,
(email: duc.ngovu@hust.edu.vn).}\thanks{T. V. Luong and N. C. Luong are with the Faculty of Computer Science,
Phenikaa University, Hanoi 12116, Vietnam (e-mail: \{thien.luongvan,
luong.nguyencong\}@phenikaa-uni.edu.vn).}\thanks{M.-T. Le is with the MobiFone R\&D Center, MobiFone Corporation, Hanoi
11312, Vietnam, (e-mail: tuan.minh@mobifone.vn).}\thanks{X.-N. Tran and T. T. H. Le are with the Advanced Wireless Communications
Group, Le Quy Don Technical University, Ha Noi 11355, Vietnam (e-mail:
\{namtx, huyen.ltt\}@mta.edu.vn).}}
\markboth{}{}
\maketitle
\begin{abstract}
This paper analyzes the bit error rate (BER) of multicarrier index
keying - orthogonal frequency division multiplexing (MCIK-OFDM) with selection combining (SC) diversity reception. Particularly, we
propose a generalized framework to derive the BER for both the low-complexity greedy detector (GD) and maximum
likelihood (ML) detector.
Based on this, closed-form expressions for the BERs of MCIK-OFDM with
the SC using either the ML or the GD are derived in presence of the
channel state information (CSI) imperfection. The asymptotic analysis
is presented to gain helpful insights into effects of different CSI conditions
on the BERs of these two detectors. More importantly, we theoretically
provide opportunities for using the GD instead of the ML under each
specific CSI uncertainty, which depend on the number of receiver antennas and the $M$-ary modulation
size. Finally, extensive simulation results
are provided in order to validate our theoretical expressions and analysis. 
\end{abstract}


\section{Introduction}

\textcolor{black}{Multicarrier index keying - orthogonal frequency division multiplexing
(MCIK-OFDM) or the so-called OFDM with index modulation (OFDM-IM)
is an emerging multicarrier scheme \cite{HassIM2009,Tsonev2011,basar3013},
which can offer higher energy efficiency and reliability over conventional
OFDM. In MCIK-OFDM, a subset of subcarriers are active to carry
data bits through both the conventional $M$-ary symbols and the indices of active subcarriers. Hence, MCIK-OFDM provides a promising
trade-off between spectral efficiency (SE) and reliability compared
to OFDM just by varying the number of active sub-carriers.}

Recently, various MCIK or IM concepts have been proposed for OFDM,
which can be found in the overview \cite{SurveyIM}. \textcolor{black}{Particularly, the IM concept was first applied to OFDM-based multicarrier modulation in \cite{HassIM2009}, and its enhanced version was proposed in \cite{Tsonev2011}, while its generalized version which independently applies the IM to different subcarrier groups was developed in \cite{basar3013}.} For the performance analysis,
in \cite{tightBound2014}, a tight bound on the bit error rate (BER)
of OFDM-IM using the maximum likelihood (ML) detection was derived.
The MCIK concept was applied to multiple input multiple output (MIMO)
systems in \cite{mimoIMbasar2016}. In \cite{GeneralizedIM}, the
generalized MCIK scheme with a variable number of active subcarriers
was proposed. In \cite{CIbasar2015}, coordinate interleaving OFDM-IM
was proposed to improve the diversity order. Also inspired by the MCIK
concept, code index modulation (CIM) as well as its generalized version were
studied in \cite{Kaddoum2015,Kaddoum2016}. Aiming to enhance the
error performance of MCIK-OFDM, several transmit diversity schemes
are reported in \cite{ThienTWC2018,codedIM2017choi,Le2020repeated,ThienTVT2018},
in which the repetition code for either the index or $M$-ary symbol
was used in \cite{ThienTWC2018,codedIM2017choi,Le2020repeated}, while
the spreading code was used in \cite{ThienTVT2018}. Meanwhile, there
are a number of studies in \cite{inPhaseQ,dualmode,multimodeIM2017}
that focus on improving the SE of MCIK-OFDM, where the IM-based transmitters
are designed to increase the number of either index or $M$-ary bits. \textcolor{black}{Recently, deep neural networks (DNNs) have been
applied to the MCIK signal detection in \cite{DeepIM2019,Wang2020Conv},
which can provide a near-optimal performance at low runtime complexity. Additionally, the use of a DNN structure called autoencoder for jointly optimizing both the transmitter and receiver of multicarrier systems was reported in \cite{Luong2020engery, Luong2021mcae,Chao2022tubro, Luong2022optical}, where the resulting learning-based systems can even achieve better error performance than IM-based multicarrier systems. Finally, the IM technique was applied to visible light communications for improving the BER performance in \cite{Khalid2021}.}

\begin{table*}[ht]
\centering \caption{Contribution Comparison of MCIK-OFDM Performance Analysis}
\label{tab:contribution} \linespread{1.0} %
\begin{tabular}{|l|c|c|c|c|c|c|}
\hline 
Contribution & \cite{basar3013} & \cite{tightBound2014} & \cite{ThienTVT2017} & \cite{thienBERGD} & \cite{JamesTVT} & This work\tabularnewline
\hline 
\hline 
BER analysis & \checkmark & \checkmark &  & \checkmark &  & \checkmark\tabularnewline
\hline 
SEP analysis &  &  & \checkmark &  & \checkmark & \tabularnewline
\hline 
Imperfect CSI & \checkmark &  & \checkmark & \checkmark &  & \checkmark\tabularnewline
\hline 
SC-based multiple-antenna receivers  &  &  &  &  & \checkmark & \checkmark\tabularnewline
\hline 
Greedy detector &  &  & \checkmark & \checkmark & \checkmark & \checkmark\tabularnewline
\hline 
ML detector & \checkmark & \checkmark & \checkmark &  &  & \checkmark\tabularnewline
\hline 
Asysmptotic analysis &  &  & \checkmark & \checkmark & \checkmark & \checkmark\tabularnewline
\hline 
Theoretical guideline for detector selection &  &  &  &  &  & \checkmark\tabularnewline
\hline 
\end{tabular}
\end{table*}

Most of the aforementioned papers consider the ML or log-likelihood
ratio (LLR) detector for MCIK-OFDM, which still has a significantly
higher complexity than the classical OFDM. In \cite{GDjamesPIMRC2015},
a low-complexity greedy detector (GD) was developed, which utilizes the
energy detection method to estimate the active indices before decoding the
$M$-ary symbols conveyed on these active sub-carriers. The outage
probabilities and the pair-wise error probability of the GD under generalized
fading were analyzed in \cite{Pout2017} and \cite{Lefteris2016},
respectively. The symbol error probability (SEP) and BER of the
GD in the presence of channel state information (CSI) imperfection
were investigated in \cite{ThienTVT2017,thienBERGD}, which reveal
that the GD detector is less sensitive to imperfect CSI than its ML counterpart. \textcolor{black}{In order to further
improve the diversity gain of GD, MCIK-OFDM
with hybrid GD and diversity receptions, namely selection combining (SC) and maximal ratio combining
(MRC), was proposed in \cite{JamesTVT} to examine the SEP,
however only for the perfect CSI case. Moreover, \cite{JamesTVT}
fails to provide an analytical comparison between the MRC/SC-based
GD and ML detectors, and its theoretical results are not tight, even at
high signal-to-noise ratios (SNRs). Hence, this work is unable to provide a theoretical guideline of selecting a suitable detection method, particularly under different CSI uncertainties. Meanwhile, the GD shown in 
\cite{ThienTVT2017,thienBERGD} is more effective in practical
systems with imperfect CSI. Therefore, it is worth investigating the
performance of MCIK-OFDM with such low-complexity MRC/SC-based GD
receivers under practical CSI uncertainty, and compare with its
ML counterpart. In addition, the performance analysis of MCIK-OFDM
using both MRC/SC and ML detection has been overlooked in the literature.}

\textcolor{black}{To address the aforementioned issues, in this paper, we first
analyze and compare the BERs of MCIK-OFDM with the SC-based multiple-antenna
receivers called MCIK-OFDM-SC, employing both ML and GD detectors, over
uncertain CSI. In particular, the main contributions of this work compared with the existing works are listed in Table~\ref{tab:contribution}, and are summarized as follows:}
\begin{itemize}
\item We propose a generalized framework for deriving the BERs of both the
GD and the ML receivers for MCIK-OFDM, where the BER is represented
as a linear combination of the
SEP and index error probability (IEP)  of the classical $M$-ary data symbols. 
\item Utilizing this proposed framework, tight, closed-form expressions for the BERs
of MCIK-OFDM-SC employing both the GD and ML detectors are derived in
presence of various CSI conditions, namely perfect CSI, and fixed
or variable CSI uncertainties. 
\item Based on the derived expressions, asymptotic results are demonstrated
to further investigate effects of different CSI uncertainties on the BERs
of the two detectors. More importantly, we asymptotically develop
conditions under which using the GD instead of the ML is desired for
MCIK-OFDM-SC under each CSI condition, particularly when the number
of antennas at the receiver increases. 
\item Simulation results are provided to validate the derived expressions, as well as theoretical guidelines for selecting detection type for each CSI condition. Unlike \cite{JamesTVT}, our theoretical results are
tight in a wide range of SNRs.
\end{itemize}
The rest of our paper is as follows. Section~II describes
 MCIK-OFDM-SC and its signal detection under uncertain CSI.
In Section~III analyzes the BERs of both ML and GD, followed
by asymptotic analysis in Section~IV. Simulation results are performed
in Section~V. Section~VI concludes our paper.

\textit{Notation:}  Lower-case bold and Upper-case bold letters and
are used for vectors and matrices, respectively.  $C\left(,\right)$ and $(.)^{T}$ denotes the binomial coefficient and transpose operation, respectively.
The floor function is represented by $\left\lfloor .\right\rfloor $. 
$\mathcal{CN}\left(0,\sigma^{2}\right)$ stands for the complex Gaussian
distribution with zero mean and variance $\sigma^{2}$. $\mathbb{E}\left\{ .\right\} $
and $\mathcal{M}\left(.\right)$ present the expectation operator
and the moment generating function (MGF), respectively.

\section{System Model}

\subsection{MCIK-OFDM-SC}

\begin{figure*}[t]
\begin{centering}
\includegraphics[width=1.6\columnwidth]{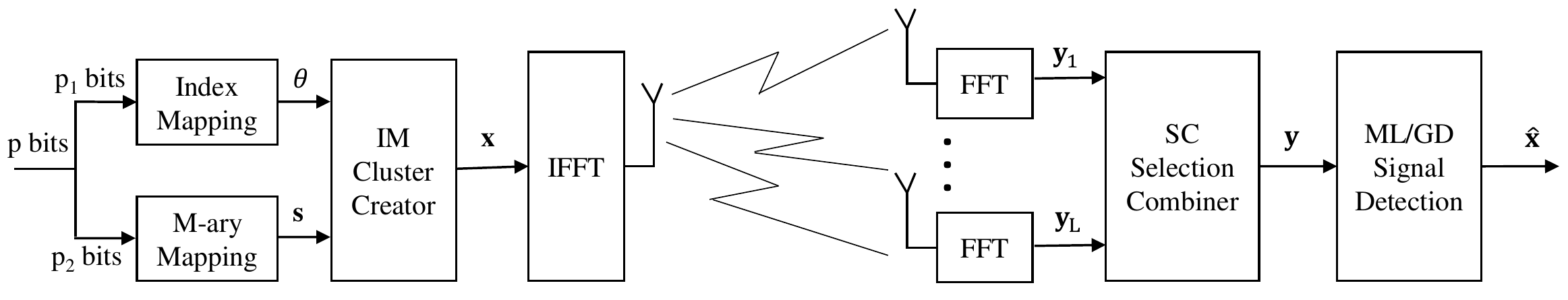}
\par\end{centering}
\caption{\textcolor{black}{The block diagram of MCIK-OFDM-SC.}\label{fig:mcik-sc}}
\end{figure*}

Consider a uplink single-input multi-output (SIMO) MCIK-OFDM scheme
with $N_{c}=NG$ sub-carriers that are divided into $G$ clusters
with $N$ sub-carriers per cluster. The transmitter employs a single antennas
while the receiver uses $L$ antennas. At the receiver, the SC technique
is employed to combine signals received from $L$ branches. Then,
the output of the SC is used to estimate transmitted data bits using either the ML or
the GD \cite{JamesTVT}. The resulting scheme is called as MCIK-OFDM-SC. \textcolor{black}{Since each cluster independently operates the MCIK-OFDM technique,
for simplicity and without loss of generality, hereinafter we address
the problem of only one cluster, whose block diagram is  illustrated in Fig.~\ref{fig:mcik-sc}. Here, the role of OFDM framework is to make sub-carriers orthogonal to each other so that we can independently apply the MCIK concept to each cluster, reducing the transceiver complexity.}

In every MCIK-OFDM transmission per cluster, only $K$ out of $N$
sub-carriers are activated to carry information bits with $K$ complex $M$-ary
symbols, while additional data bits are delivered by the indices of
active sub-carriers. More specifically, $p$ incoming bits are partitioned
into two streams ($p=p_{1}+p_{2}$) at the transmitter. Utilizing
combinatorial method or look-up table (LUT) \cite{basar3013}, the
first $p_{1}$ bits are mapped to a pattern of $K$ active sub-carriers.
Denote by $\theta=\left\{ \alpha_{1},...,\alpha_{K}\right\} $ the
set of $K$ active sub-carrier indices, where $\alpha_{k}\in\left\{ 1,...,N\right\} $
for $k=1,...,K$. Note that $\theta$ can be referred to as an index
symbol, which is identified by $p_{1}$ index bits. The remaining
$p_{2}$ bits are mapped to $K$ $M$-ary symbols. For given
$N,K$ and $M$, the number of index bits and symbol bits are given
by $p_{1}=\left\lfloor \log_{2}C\left(N,K\right)\right\rfloor $ and
$p_{2}=K\log_{2}M$, respectively. Denote by $\mathcal{S}$ the $M$-ary
constellation. Using $\theta$ and $K$ non-zero symbols (determined
by $p$ incoming bits), the transmitted signal for each cluster is
given as $\mathbf{x}=\left[x\left(1\right),...,x\left(N\right)\right]^{T},$
where $x\left(\alpha\right)=0$ for $\alpha\notin\theta$ and $x\left(\alpha\right)\in\mathcal{S}$
for $\alpha\in\theta$. \textcolor{black}{Here, note that the $K$ non-zero data symbols conveyed on active sub-carriers are denoted as vector $\mathbf{s}$ in Fig.~\ref{fig:mcik-sc}. The frequency domain signal $\mathbf{x}$ is then processed by the inverse fast Fourier transform (IFFT) before being transmitted to the receiver.}

\textcolor{black}{The received signal at the $l$-th antenna in the frequency domain, i.e., the signal obtained after the FFT,
is given by} 
\begin{equation}
\mathbf{y}_{l}=\mathbf{H}_{l}\mathbf{x}+\mathbf{n}_{l},
\end{equation}
where $\mathbf{H}_{l}=\text{diag}\left\{ h_{l}\left(1\right),...,h_{l}\left(N\right)\right\} $
is the channel matrix between the transmitter and the $l$-th receiver
antenna, while $\mathbf{n}_{l}=\left[n_{l}\left(1\right),...,n_{l}\left(N\right)\right]^{T}$
is the noise vector with $n_{l}\left(\alpha\right)\sim\mathcal{CN}\left(0,N_{0}\right)$,
for $\alpha=1,...,N$ and $l=1,...,L$. \textcolor{black}{Particularly, $h_{l}\left(\alpha\right)$ represents the Rayleigh fading
channel, which is identical and independent to each other, where  $h_{l}\left(\alpha\right)\sim\mathcal{CN}\left(0,1\right)$. Here, we assume that the cyclic prefix inserted to each OFDM symbol in the time domain is large enough to completely combat the inter-symbol interference \cite{basar3013}.} As such, the average SNR per active
sub-carrier is given by $\bar{\gamma}=\varphi E_{s}/N_{0},$ where
$E_{s}$ denotes the average power per non-zero $M$-ary symbol and
$\varphi=N/K$ is the power allocation ratio.


\subsection{Post-Combining Detection under CSI Uncertainty}

We consider a practical MCIK-OFDM-SC system where the receiver imperfectly
knows the CSI. Particularly, denote by $\hat{h}_{l}\left(\alpha\right)$
the estimate of the true channel $h_{l}\left(\alpha\right)$, and
we have 
\begin{equation}
\hat{h}_{l}\left(\alpha\right)=h_{l}\left(\alpha\right)-e_{l}\left(\alpha\right),\label{eq:h_hat}
\end{equation}
where $e_{l}\left(\alpha\right)$ represents the channel estimation
error as being independent of $\hat{h}_{l}\left(\alpha\right)$ and
$e_{l}\left(\alpha\right)\sim\mathcal{CN}\left(0,\epsilon^{2}\right)$,
and $\hat{h}_{l}\left(\alpha\right)\sim\mathcal{CN}\left(0,1-\epsilon^{2}\right),$
where $\epsilon^{2}\in\left[0,1\right)$ denotes the error variance.

For sub-carrier $\alpha$, the $l^{*}$-th branch is selected as the
output of the SC such that $l^{*}=\arg\max_{l}\left|\hat{h}_{l}\left(\alpha\right)\right|^{2}$.
Hence, the output signal of the SC can be given by 
\begin{equation}
\mathbf{y}=\mathbf{H}\mathbf{x}+\mathbf{n},\label{eq:y_SC}
\end{equation}
where $\mathbf{H}=\text{diag}\left\{ h\left(1\right),...,h\left(N\right)\right\} $
denotes the channel matrix of the SC and the corresponding noise vector
is $\mathbf{n}=\left[n\left(1\right),...,n\left(N\right)\right]^{T}$,
where $h\left(\alpha\right)=h_{l^{*}}\left(\alpha\right)$ and $n\left(\alpha\right)=n_{l^{*}}\left(\alpha\right)$
for $\alpha=1,...,N$. Notice in \eqref{eq:y_SC} that $\mathbf{y}=\left[y\left(1\right),...,y\left(N\right)\right]^{T},$
with $y\left(\alpha\right)=h\left(\alpha\right)x\left(\alpha\right)+n\left(\alpha\right)$.

Let $\hat{h}\left(\alpha\right)$ be the estimate of $h\left(\alpha\right)$,
i.e., $\hat{h}\left(\alpha\right)=\hat{h}_{l^{*}}\left(\alpha\right)$.
Based on $\mathbf{y}$ and $\hat{h}\left(\alpha\right),$ either the
ML or the GD can be employed for the signal detection as follows.

\subsubsection{Post-Combining ML }

Under imperfect CSI, the estimated signal $\hat{\mathbf{x}}$ is calculated
by the ML criterion as 
\[
\hat{\mathbf{x}}=\arg\min_{\mathbf{x}}\left\Vert \mathbf{y}-\mathbf{\hat{H}x}\right\Vert ^{2},
\]
where $\hat{\mathbf{H}}=\text{diag}\left\{ \hat{h}\left(1\right),...,\hat{h}\left(N\right)\right\} $
denotes the estimate of the channel matrix after the SC. Utilizing
$\hat{\mathbf{x}}$, the index symbol $\hat{\theta}$ and $K$ non-zero
symbols $x\left(\alpha\right)$ with $\alpha\in\hat{\theta}$ are
recovered.

\subsubsection{Post-Combining GD }

Post-combining GD makes best antenna selections per sub-carrier before
GD processing. For given $\mathbf{H}$, the GD detects signals through
two following steps. Firstly, the active indices are estimated by
$K$ sub-carriers that have the largest SC-output energies, i.e.,
$\left|y\left(\alpha\right)\right|^{2}.$ Secondly, the non-zero $M$-ary
symbols are detected by applying the ML decision to activated sub-carrier
$\alpha$ as

\begin{equation}
x\left(\alpha\right)=\arg\min_{x\left(\alpha\right)\in\mathcal{S}}\left|y\left(\alpha\right)-\hat{h}\left(\alpha\right)x\left(\alpha\right)\right|^{2}.\label{eq:step2_GD}
\end{equation}

Note that the GD detector has not only lower complexity, but also less sensitivity
to CSI imperfection, than the ML detector \cite{thienBERGD}. However, when
the number of antennas is limited to one, the ML still
perform much better than GD under certain CSI conditions, especially when $M$
is small (e.g., $M=2,4$) \cite{ThienTVT2017}.

As a result, we are prompted to examine the BER performance of both
the GD and the ML in MCIK-OFDM-SC, in order to understand if the post-combining
GD receiver benefits from diversity gain. For this, we intend to derive
the closed-form expressions for the BERs of the two detectors, taking CSI uncertainty into consideration in the next section.

\section{BER Analysis With CSI Uncertainty}

We note that the ML performs the same performance as the log-likelihood
ratio (LLR) detector \cite{codedIM2017choi} which also has two separate
steps as the GD. Thus, we now introduce a generalized framework to
derive the BERs of both the ML and the GD. Particularly, we consider
bit error event consisting of two parts: the index bit error ($p_{1}$
bits) and the symbol bit error ($p_{2}$ bits). Let $P_{1}$ be the
index BER (IBER) and $P_{2}$ be the symbol BER (SBER). Then, the
BER of either the ML or the GD is given by 
\begin{equation}
P_{b}=\frac{p_{1}P_{1}+p_{2}P_{2}}{p_{1}+p_{2}}.\label{eq:Pb_start}
\end{equation}
The IBER and the SBER are obtained by \cite{thienBERGD} 
\begin{equation}
P_{1}\approx\eta\overline{P}_{I}/2,\label{eq:P1}
\end{equation}
\begin{equation}
P_{2}\le\frac{\overline{P}_{I}}{2K}+\frac{\overline{P}_{M}}{\log_{2}M},\label{eq:P2}
\end{equation}
where $\overline{P}_{I}$ denotes the average index error probability
(IEP), $\eta=1$ for $N>2$ and $\eta=2$ for $N=2$, and $\overline{P}_{M}$
is the average SEP of the $M$-ary symbol detection as long as
the activated indices are correctly detected. Plugging \eqref{eq:P1}
and \eqref{eq:P2} into \eqref{eq:Pb_start}, the generalized BER
expression for both the ML and the GD is given by
\begin{equation}
P_{b}\approx\frac{\left(\eta p_{1}+m\right)\overline{P}_{I}/2+K\overline{P}_{M}}{p},\label{eq:Pb_general}
\end{equation}
where $m=\log_{2}M$ and $p=p_{1}+p_{2}$.

\textit{Remark 1:} As seen from \eqref{eq:Pb_general}, when $K$
increases to $N$, the BER of either the ML or the GD approaches that
of classical OFDM, which is $\overline{P}_{M}/m$. As a result, the
performance gap between these two detectors gets smaller when $K$
gets larger.

\textit{Remark 2:} $\overline{P}_{M}$ in \eqref{eq:Pb_general} is
the same for both the ML and the GD, while $\overline{P}_{I}$ depends
on the detection type employed. Thus, to find out the BER expressions for
the GD and the ML in MCIK-OFDM-SC, we need to derive $\overline{P}_{I}$
for them, considering CSI uncertainty. Meanwhile, $\overline{P}_{M}$
is provided in the following lemma when the $M$-ary PSK modulation is
employed.
\begin{lem}
Under CSI uncertainty with the error variance $\epsilon^{2}$, the
average SEP of the conventional $M$-ary PSK symbol detection in MCIK-OFDM-SC
is approximated by 
\begin{equation}
\overline{P}_{M}\approx\frac{\xi}{12}\left\{ \frac{L!}{\prod_{l=1}^{L}\left[l+\frac{\left(1-\epsilon^{2}\right)\rho\bar{\gamma}}{1+\epsilon^{2}\bar{\gamma}}\right]}+\frac{3L!}{\prod_{l=1}^{L}\left[l+\frac{4\left(1-\epsilon^{2}\right)\rho\bar{\gamma}}{3\left(1+\epsilon^{2}\bar{\gamma}\right)}\right]}\right\} ,\label{eq:PM}
\end{equation}
where $\rho=\sin^{2}\left(\pi/M\right)$, $\xi=1$ for $M=2$ and
$\xi=2$ for $M>2$. 
\end{lem}
\begin{IEEEproof}
See Appendix A. 
\end{IEEEproof}

\subsection{BER for ML with SC Reception and CSI Uncertainty}

We first consider the IEP of the ML in MCIK-OFDM with the SC and imperfect
CSI. Denote by $P_{I_{1}}$ the instantaneous IEP of the ML, which
is approximated by \cite{ThienTVT2017} 
\begin{equation}
P_{I_{1}}\approx\frac{K}{N}\sum_{\alpha=1}^{N}\sum_{\tilde{\alpha}\ne\alpha=1}^{N-K}\left[\frac{1}{12}e^{-\frac{\bar{\gamma}\left(\hat{\nu}_{\alpha}+\hat{\nu}_{\tilde{\alpha}}\right)}{4+2\bar{\gamma}\epsilon^{2}}}+\frac{1}{4}e^{-\frac{2\bar{\gamma}\left(\hat{\nu}_{\alpha}+\hat{\nu}_{\tilde{\alpha}}\right)}{6+3\bar{\gamma}\epsilon^{2}}}\right],\label{eq:PI_1}
\end{equation}
where $\hat{\nu}_{\alpha}=\left|\hat{h}\left(\alpha\right)\right|^{2}$,
$\hat{\nu}_{\tilde{\alpha}}=\left|\hat{h}\left(\tilde{\alpha}\right)\right|^{2}$.

Denote $\hat{\nu}_{\varSigma}=\hat{\nu}_{\alpha}+\hat{\nu}_{\tilde{\alpha}}$.
The moment generating function (MGF) of $\hat{\nu}_{\varSigma}$ can
be attained by $\mathcal{M}_{\hat{\nu}_{\varSigma}}\left(s\right)=\mathcal{M}_{\hat{\nu}}^{2}\left(s\right),$
where $\mathcal{M}_{\hat{\nu}}\left(s\right)$ is the MGF of $\hat{\nu}_{\alpha}$
which is given in \eqref{eq:MGF_v_hat}. Here, applying the MGF approach
to \eqref{eq:PI_1}, we obtain the average IEP of the ML with the
SC and uncertain CSI as follows 
\begin{equation}
\overline{P}_{I_{1}}\approx\frac{\Psi_{1}}{12}\left\{ \frac{\left(L!\right)^{2}}{\prod_{l=1}^{L}\left[l+\frac{\left(1-\epsilon^{2}\right)\bar{\gamma}}{4+2\bar{\gamma}\epsilon^{2}}\right]^{2}}+\frac{3\left(L!\right)^{2}}{\prod_{l=1}^{L}\left[l+\frac{2\left(1-\epsilon^{2}\right)\bar{\gamma}}{6+3\bar{\gamma}\epsilon^{2}}\right]^{2}}\right\} ,\label{eq:PI_1_ave}
\end{equation}
where $\Psi_{1}=K\left(N-K\right)$.

As observed from \eqref{eq:PI_1_ave}, note that as $L=1$, the
average IEP of the ML in \eqref{eq:PI_1_ave} reduces to \cite[Eq. (16)]{ThienTVT2017}.
In addition, $\overline{P}_{I_{1}}$ mainly relies on $N,K$ and $L$,
while being less influenced by the modulation size $M$.

Finally, the BER of the ML (denoted by $P_{b_{1}}$) can be obtained
by inserting \eqref{eq:PM} and \eqref{eq:PI_1_ave} to \eqref{eq:Pb_general}
as 
\begin{align}
P_{b_{1}} & \approx\frac{\widetilde{\Psi}_{1}}{24p}\left\{ \frac{\left(L!\right)^{2}}{\prod_{l=1}^{L}\left[l+\frac{\left(1-\epsilon^{2}\right)\bar{\gamma}}{4+2\bar{\gamma}\epsilon^{2}}\right]^{2}}+\frac{3\left(L!\right)^{2}}{\prod_{l=1}^{L}\left[l+\frac{2\left(1-\epsilon^{2}\right)\bar{\gamma}}{6+3\bar{\gamma}\epsilon^{2}}\right]^{2}}\right\} \nonumber \\
 & +\frac{K\xi}{12p}\left\{ \frac{L!}{\prod_{l=1}^{L}\left[l+\frac{\left(1-\epsilon^{2}\right)\rho\bar{\gamma}}{1+\epsilon^{2}\bar{\gamma}}\right]}+\frac{3L!}{\prod_{l=1}^{L}\left[l+\frac{4\left(1-\epsilon^{2}\right)\rho\bar{\gamma}}{3\left(1+\epsilon^{2}\bar{\gamma}\right)}\right]}\right\} ,\label{eq:Pb_1}
\end{align}
where $\widetilde{\Psi}_{1}=\Psi_{1}\left(\eta p_{1}+m\right)=K\left(N-K\right)\left(\eta p_{1}+m\right).$

It is shown from \eqref{eq:Pb_1} that increasing $L$ improves the
BER of the ML. Moreover, for given $N,$ $L$ and $\bar{\gamma}$,
the BER $P_{b_{1}}$ depends on both $K$ and $\epsilon^{2}$. For
example, when $K$ gets larger, the second term, which is related to the $M$-ary
symbol detection, will dominate over $P_{b_{1}}$. Especially, as $K=N$,
\eqref{eq:Pb_1} reduces to the BER of the classical OFDM.

\subsection{BER for GD with SC Reception and CSI Uncertainty}

In MCIK-OFDM with the single antenna used at both the transmitter
and the receiver, the IEP of the GD is independent of CSI conditions
\cite{thienBERGD}. However, this is no longer true when employing
the SC for MCIK-OFDM. Particularly, the instantaneous IEP of the GD
is given by \cite{thienBERGD,JamesTVT} 
\begin{equation}
P_{I_{2}}=\frac{K}{N}\sum_{\alpha=1}^{N}\sum_{i=1}^{N-K}\frac{\left(-1\right)^{i+1}C\left(N-K,i\right)}{i+1}e^{-\frac{i\bar{\gamma}\nu_{\alpha}}{i+1}},\label{eq:PI_start}
\end{equation}
where $\nu_{\alpha}=\left|h\left(\alpha\right)\right|^{2}$ which
is obviously affected by the estimate $\hat{h}_{l}\left(\alpha\right)$
due to $h\left(\alpha\right)=h_{l^{*}}\left(\alpha\right)$ with $l^{*}=\max_{l}\left|\hat{h}_{l}\left(\alpha\right)\right|^{2}$. \textcolor{black}{The detailed derivation of \eqref{eq:PI_start} over Rayleigh fading channels was presented in \cite{GDjamesPIMRC2015}, which is not included here for the sake of brevity.} 
Thus, the IEP of the GD in our system depends on the channel estimation
errors. This makes the derivation of the average IEP for this detector
non-trivial as follows.

First, it is needed to figure out the MGF of $\nu_{\alpha}$. Using
\eqref{eq:h_hat}, $h\left(\alpha\right)$ can be represented as $h\left(\alpha\right)=e^{j\phi}\left|\hat{h}\left(\alpha\right)\right|+e\left(\alpha\right)=e^{j\phi}\left(\left|\hat{h}\left(\alpha\right)\right|+\tilde{e}\left(\alpha\right)\right)$,
where $\tilde{e}\left(\alpha\right)=e^{-j\phi}e\left(\alpha\right)\sim\mathcal{CN}\left(0,\epsilon^{2}\right)$
and $\phi$ denotes the argument of $\hat{h}\left(\alpha\right)$.
This results in 
\begin{equation}
\left|h\left(\alpha\right)\right|^{2}=\left|\left|\hat{h}\left(\alpha\right)\right|+\tilde{e}\left(\alpha\right)\right|^{2}.\label{eq:v_alp}
\end{equation}
From \eqref{eq:v_alp}, the MGF of $\nu_{\alpha}$ can be computed
as 
\begin{align}
\mathcal{M}_{\nu}\left(t\right) & =\mathbb{E}_{\left|h\left(\alpha\right)\right|^{2}}\left\{ e^{\left|h\left(\alpha\right)\right|^{2}t}\right\} \nonumber \\
 & =\mathbb{E}_{\left|\hat{h}\left(\alpha\right)\right|^{2}}\left\{ \mathbb{E}_{\left|\left|\hat{h}\left(\alpha\right)\right|+\tilde{e}\left(\alpha\right)\right|^{2}}\left\{ e^{\left|\left|\hat{h}\left(\alpha\right)\right|+\tilde{e}\left(\alpha\right)\right|^{2}t}\right\} \right\} \nonumber \\
 & =\int_{0}^{\infty}f_{\left|\hat{h}\left(\alpha\right)\right|^{2}}\left(x\right)\mathcal{M}_{\left|\left|\hat{h}\left(\alpha\right)\right|+\tilde{e}\left(\alpha\right)\right|^{2}}\left(t\right)dx,\label{eq:MGF_v_start}
\end{align}
which motivates us to propose the following lemma. 
\begin{lem}
Let $\tilde{e}\left(\alpha\right)\sim\mathcal{CN}\left(0,\epsilon^{2}\right),$
then for given $\left|\hat{h}\left(\alpha\right)\right|^{2},$ the
MGF of $\left|\left|\hat{h}\left(\alpha\right)\right|+\tilde{e}\left(\alpha\right)\right|^{2}$
is given by 
\begin{equation}
\mathcal{M}_{\left|\left|\hat{h}\left(\alpha\right)\right|+\tilde{e}\left(\alpha\right)\right|^{2}}\left(t\right)=\frac{e^{\frac{\left|\hat{h}\left(\alpha\right)\right|^{2}t}{1-\epsilon^{2}t}}}{1-\epsilon^{2}t}.\label{eq:MGF_non_central}
\end{equation}
\end{lem}
\begin{IEEEproof}
See Appendix B. 
\end{IEEEproof}
Inserting \eqref{eq:PDF_v_hat} and \eqref{eq:MGF_non_central} into
\eqref{eq:MGF_v_start}, through simple manipulations, we obtain 
\begin{equation}
\mathcal{M}_{\nu}\left(t\right)=\frac{L!}{\left(1-\epsilon^{2}t\right)\prod_{l=1}^{L}\left[l-\frac{\left(1-\epsilon^{2}\right)t}{1-\epsilon^{2}t}\right]}.\label{eq:MGF_v_final}
\end{equation}

Note that to the best of our knowledge, the approach to derive the
MGF of $\nu_{\alpha}$ in closed-form \eqref{eq:MGF_v_final} is novel.
This interestingly results in a simple, exact closed-form expression
for the average IEP of the GD with the SC and uncertain CSI, by applying
the MGF approach to \eqref{eq:PI_start} and using \eqref{eq:MGF_v_final},
as 
\begin{equation}
\overline{P}_{I_{1}}=K\sum_{i=1}^{N-K}\frac{\left(-1\right)^{i+1}C\left(N-K,i\right)L!}{\left(i+1+i\epsilon^{2}\bar{\gamma}\right)\prod_{l=1}^{L}\left[l+\frac{\left(1-\epsilon^{2}\right)i\bar{\gamma}}{i+1+i\epsilon^{2}\bar{\gamma}}\right]}.\label{eq:PI_ave}
\end{equation}

As observed from \eqref{eq:PI_ave}, when $L=1$, the expression for
$\overline{P}_{I}$ becomes \cite[Eq. (8)]{thienBERGD} which no longer
depends on $\epsilon^{2}.$ In addition, as $L>1$, the IEP performance
suffers from a degradation caused by CSI uncertainty, i.e., $\epsilon^{2}.$
Note that for any $\epsilon^{2}\in\left[0,1\right)$, $\overline{P}_{I_{1}}$
always tends to 0 as $\bar{\gamma}$ increases to infinity, even for
the worst case of $\epsilon^{2}=1$.

Finally, the BER of the GD with the SC and uncertain CSI can be attained 
by substituting \eqref{eq:PM} and \eqref{eq:PI_ave} to \eqref{eq:Pb_general}
as follows:
\begin{align}
P_{b_{2}} & \approx\frac{K\left(\eta p_{1}+m\right)}{2p}\sum_{i=1}^{N-K}\frac{\left(-1\right)^{i+1}C\left(N-K,i\right)L!}{\left(i+1+i\epsilon^{2}\bar{\gamma}\right)\prod_{l=1}^{L}\left[l+\frac{\left(1-\epsilon^{2}\right)i\bar{\gamma}}{i+1+i\epsilon^{2}\bar{\gamma}}\right]}\nonumber \\
 & +\frac{K\xi}{12p}\left\{ \frac{L!}{\prod_{l=1}^{L}\left[l+\frac{\left(1-\epsilon^{2}\right)\rho\bar{\gamma}}{1+\epsilon^{2}\bar{\gamma}}\right]}+\frac{3L!}{\prod_{l=1}^{L}\left[l+\frac{4\left(1-\epsilon^{2}\right)\rho\bar{\gamma}}{3\left(1+\epsilon^{2}\bar{\gamma}\right)}\right]}\right\} .\label{eq:Pb_2}
\end{align}

Observe from \eqref{eq:Pb_2} that different from MCIK-OFDM with the
single antenna \cite{thienBERGD}, where $\epsilon^{2}$ affects only
the term related to the $M$-ary symbol detection, in MCIK-OFDM-SC
having multiple anttenas, $\epsilon^{2}$ influences on both the index
detection error and the $M$-ary symbol detection error. As $L=1$,
\eqref{eq:Pb_2} reduces to \cite[Eq. (15)]{thienBERGD}, which confirms
the accuracy of our derivation for the BER expression of MCIK-OFDM-SC.

\section{Asymptotic Analysis}

We now carry out the asymptotic analysis for the BERs of both ML
and GD detectors at high SNRs and in a large number of antennas. In particular,
we investigate the impact of various CSI uncertainties, namely perfect CSI,
fixed CSI uncertainty, and minimum mean square (MMSE) based variable CSI uncertainty. In addition,
the performance comparison between the two detectors is provided.
This allows to recommend that when the GD should be used under each
CSI condition as the number of antennas increases.

Note that existing studies \cite{JamesTVT,ThienTVT2017,thienBERGD}
have not provided any analytical comparisons between the ML and the
GD such as the behavior of the coding gain gap between them when the
number of antennas changes. Moreover, \cite{JamesTVT} even has not
included any asymptotic analysis for the GD with the SC. 

\subsection{Perfect CSI $(\epsilon^{2}=0)$}

As $\epsilon^{2}=0$ and $\bar{\gamma}$ tends to infinity, the BERs
in \eqref{eq:Pb_1} and \eqref{eq:Pb_2} can be asymptotically approximated
by 
\begin{equation}
P_{b_{1}}\approx\Upsilon\left(\frac{\xi\Omega}{6\rho^{L}}\right)\frac{1}{\gamma_{0}^{L}},\label{eq:Pb_1_perfect}
\end{equation}
\begin{equation}
P_{b_{2}}\approx\Upsilon\left[\left(\eta p_{1}+m\right)\omega+\frac{\xi\Omega}{6\rho^{L}}\right]\frac{1}{\gamma_{0}^{L}},\label{eq:Pb_2_perfect}
\end{equation}
where $\Upsilon=K^{L+1}L!/2pN^{L}$, $\Omega=1+3^{L+1}/4^{L}$, $\omega=\sum_{i=1}^{N-K}\left(-1\right)^{i+1}C\left(N-K,i\right)\left(1+i\right)^{L-1}/i^{L}$,
and $\gamma_{0}=E_{s}/N_{0}$ is the average SNR per sub-carrier.

As observed from \eqref{eq:Pb_1_perfect} and \eqref{eq:Pb_2_perfect},
both the ML and the GD attain a diversity order of $L$ under perfect
CSI. Moreover, for given $N$ and $L$, a smaller $K$ provides lower
BERs.

Regarding the comparison between the GD and the ML, we consider the
coding gain attained by the ML over the GD under perfect CSI (denoted
by $\Delta_{1}$), which can be denoted by $\Delta_{1}=10\log_{10}\left(P_{b_{2}}/P_{b_{1}}\right)^{1/L}$.
Using \eqref{eq:Pb_1_perfect} and \eqref{eq:Pb_2_perfect}, we have
\begin{equation}
\Delta_{1}=\frac{10}{L}\log_{10}\left(1+\eta_{1}\right)\text{\,(dB)},\label{eq:Delta_1}
\end{equation}
where $\eta_{1}=6\left(\eta p_{1}+m\right)\omega\rho^{L}/\xi\Omega$.
Based on this result, we introduce the following theorem.

\textbf{Theorem 1.} \textit{Consider MCIK-OFDM with the SC and perfect
CSI. For $M=2$, the ML performs better than the GD in terms of the BER by
3 dB, at large $L$, i.e, $\lim_{L\rightarrow\infty}\Delta_{1}\approx3$
(dB). For $M\ge4$, the BER of GD approaches to that of ML
when increasing $L$, i.e., $\lim_{L\rightarrow\infty}\Delta_{1}=0$
(dB). Especially, when $M\ge8$, the BERs of the two detectors rapidly
converge to each other as $L$ increases, i.e., $\lim_{L\rightarrow\infty}\eta_{1}=0$.} 
\begin{IEEEproof}
Since $\omega$ in \eqref{eq:Pb_2_perfect} can be approximated by
$\omega\approx\left(N-K\right)2^{L-1}$ at large $L$, we approximate
$\eta_{1}$ at large $L$ as 
\begin{equation}
\eta_{1}\approx\beta_{1}\lambda_{1}^{L},\label{eq:eta_1}
\end{equation}
where $\lambda_{1}=2\rho,$ recalling $\rho=\sin^{2}\left(\pi/M\right),$
and $\beta_{1}=3\left(\eta p_{1}+m\right)\left(N-K\right)/\xi\Omega$
which decreases when increasing $L$ due to $\Omega=1+3^{L+1}/4^{L}.$

For $M=2$, we obtain $\lambda_{1}=2$, thus $\eta_{1}\approx\beta_{1}2^{L}$.
Using \eqref{eq:Delta_1}, $\lim_{L\rightarrow\infty}\Delta_{1}=\lim_{L\rightarrow\infty}\left(10/L\right)\log_{10}\left(1+\beta_{1}2^{L}\right)=10\log_{10}2\approx3$
(dB).

For $M\ge4,$ we obtain $\lambda_{1}\le1$, thus $1<\eta_{1}\le1+\beta_{1}$.
This leads to $\lim_{L\rightarrow\infty}\Delta_{1}=0$ (dB).

For $M\ge8,$ we attain $\lambda_{1}\le2\sin^{2}\left(\pi/8\right)<0.3$,
which results in $\lim_{L\rightarrow\infty}\eta_{1}=\lim_{L\rightarrow\infty}\beta_{1}\lambda_{1}^{L}=0$. 
\end{IEEEproof}
\textit{Remark 3.} From Theorem~1, it is recommended that the GD
should be used rather than the ML under perfect CSI as $M\ge8$, especially
when $L$ gets larger. This is because the GD can achieve a nearly
optimal BER at a significantly lower complexity than the ML detector for large $M$
and $L$. Note that the complexities of the ML and GD in MCIK-OFDM
with the SC are $\mathcal{C}_{ML-SC}=N+2CM^{K}$ and $\mathcal{C}_{GD-SC}=2N+2KM,$
respectively, where $C=2^{p_{1}}$ \cite{JamesTVT}. Obviously, when
$K$ and $M$ become larger, we attain $\mathcal{C}_{ML-SC}\gg\mathcal{C}_{GD-SC}$.

\subsection{Fixed CSI Uncertainty $(\epsilon^{2}>0)$}

As $\epsilon^{2}>0$ is fixed, the BERs in \eqref{eq:PI_ave} and
\eqref{eq:Pb_2} can be rewritten at high SNRs, respectively, as follows:
\begin{align}
P_{b_{1}} & \approx\underbrace{\frac{\widetilde{\Psi}_{1}}{24p}\left\{ \frac{1}{\prod_{l=1}^{L}\left[1+\frac{\left(1-\epsilon^{2}\right)}{2l\epsilon^{2}}\right]^{2}}+\frac{3}{\prod_{l=1}^{L}\left[1+\frac{2\left(1-\epsilon^{2}\right)}{3l\epsilon^{2}}\right]^{2}}\right\} }_{A_{1}}\nonumber \\
 & +\frac{K\xi}{12p}\left\{ \frac{1}{\prod_{l=1}^{L}\left[1+\frac{\left(1-\epsilon^{2}\right)\rho}{l\epsilon^{2}}\right]}+\frac{3}{\prod_{l=1}^{L}\left[1+\frac{4\left(1-\epsilon^{2}\right)\rho}{3l\epsilon^{2}}\right]}\right\} ,\label{eq:Pb_1_fixed}
\end{align}
\begin{equation}
P_{b_{2}}\approx\frac{K\xi}{12p}\left\{ \frac{1}{\prod_{l=1}^{L}\left[1+\frac{\left(1-\epsilon^{2}\right)\rho}{l\epsilon^{2}}\right]}+\frac{3}{\prod_{l=1}^{L}\left[1+\frac{4\left(1-\epsilon^{2}\right)\rho}{3l\epsilon^{2}}\right]}\right\} ,\label{eq:Pb_2_fixed}
\end{equation}
where we recall that $\widetilde{\Psi}_{1}=K\left(N-K\right)\left(\eta p_{1}+m\right).$

As seen from \eqref{eq:Pb_1_fixed} and \eqref{eq:Pb_2_fixed}, for
fixed $\epsilon^{2}$, there exists error floors on the BERs of both
the ML and the GD, or equivalently, increasing the SNR does not improve
the BER. Thus, these two detectors in this case achieve a zero diversity
gain for any $L$. Furthermore, when $L$ gets larger or $\epsilon^{2}$
gets smaller, the error floors in \eqref{eq:Pb_1_fixed} and \eqref{eq:Pb_2_fixed}
become lower.

The following theorem compares the BER between the ML and the GD in
MCIK-OFDM with the SC and fixed $\epsilon^{2}$.

\textbf{Theorem 2.}\textit{ In MCIK-OFDM using the SC under fixed CSI
uncertainty, the GD achieves a better BER than the ML detector at high SNRs,}
i.e., $P_{b_{1}}>P_{b_{2}}$. 
\begin{IEEEproof}
It is shown from \eqref{eq:Pb_1_fixed} and \eqref{eq:Pb_2_fixed}
that at high SNRs, $P_{b_{1}}=P_{b_{2}}+A_{1}>P_{b_{2}},$ where the
term $A_{1}$ is related to the index detection error of the ML. This
concludes the proof. 
\end{IEEEproof}
\textit{Remark 4.} As a result of Theorem~2, under fixed CSI imperfection,
the GD is able to outperform the ML in terms of both the BER and 
computational complexity, even for any $M$. This is obviously contrary
to the perfect CSI case, where the BER of  ML is always lower
than that of GD.

\subsection{MMSE-Based Variable CSI Uncertainty}

Note that the error variance provided by the MMSE channel estimator
is given by \cite{thienBERGD} 
\begin{equation}
\epsilon^{2}=\frac{1}{1+\gamma_{0}},\label{eq:mmse}
\end{equation}
which varies as a decreasing function of the SNR$.$

Inserting \eqref{eq:mmse} to \eqref{eq:Pb_1} and \eqref{eq:Pb_2},
we obtain the asymptotic BERs for the ML and the GD in this case as
\begin{equation}
P_{b_{1}}\approx\Upsilon\left[\frac{\xi\Omega\left(1+N/K\right)^{L}}{6\rho^{L}}\right]\frac{1}{\gamma_{0}^{L}},\label{eq:Pb_1_MMSE}
\end{equation}
\begin{equation}
P_{b_{2}}\approx\Upsilon\left[\psi\left(\eta p_{1}+m\right)+\frac{\xi\Omega\left(1+N/K\right)^{L}}{6\rho^{L}}\right]\frac{1}{\gamma_{0}^{L}},\label{eq:Pb_2_MMSE}
\end{equation}
where $\Upsilon$ and $\Omega$ are defined in \eqref{eq:Pb_2_perfect},
and $\psi=\sum_{i=1}^{N-K}\left(-1\right)^{i+1}C\left(N-K,i\right)\left(i+1+iN/K\right)^{L-1}/i^{L}$

As seen from \eqref{eq:Pb_1_MMSE} and \eqref{eq:Pb_2_MMSE},
both the GD and the ML of MCIK-OFDM with the SC achieves the same
diversity order of $L$ in this case. However, due to the impact of
MMSE channel estimation errors, the BERs in \eqref{eq:Pb_1_MMSE}
and \eqref{eq:Pb_2_MMSE} are obviously greater than that of the perfect
CSI case. For example, we can see from \eqref{eq:Pb_1_perfect} and
\eqref{eq:Pb_1_MMSE} that under the MMSE CSI imperfection, the ML
endures a coding gain loss of $10\log_{10}\left(1+N/K\right)$
(dB) compared with the perfect CSI case.

As for the comparison in the BER between the ML and the GD, denote
by $\Delta_{2}$ the coding gain attained by the ML over GD detector under
MMSE variable CSI uncertainty, which can be obtained from \eqref{eq:Pb_1_MMSE}
and \eqref{eq:Pb_2_MMSE} as 
\begin{equation}
\Delta_{2}=\frac{10}{L}\log_{10}\left(1+\eta_{2}\right)\text{\,(dB)},\label{eq:Delta_2}
\end{equation}
where $\eta_{2}=6\psi\left(\eta p_{1}+m\right)\rho^{L}/\xi\Omega\left(1+N/K\right)^{L}.$
Similar to Theorem~1, utilizing \eqref{eq:Delta_2} we propose the
following theorem.

\textbf{Theorem 3.} \textit{Consider MCIK-OFDM using the SC and the
MMSE-based variable CSI imperfection. For $M\ge4$, the BERs of the
ML and the GD rapidly converge to each other as increasing }$L$,
i.e., $\lim_{L\rightarrow\infty}\eta_{2}=0$. \textit{When $M=2$,
the ML performs better than the GD in terms of the BER by a coding gain of
$10\log_{10}\left[1+K/\left(N+K\right)\right](dB)$, at large $L$,
i.e., $\lim_{L\rightarrow\infty}\Delta_{2}=10\log_{10}\left[1+K/\left(N+K\right)\right]$
(dB), moreover $\lim_{L\rightarrow\infty}\Delta_{2}<\lim_{L\rightarrow\infty}\Delta_{1}.$} 
\begin{IEEEproof}
Akin to Theorem~1, at large $L$, $\psi$ in \eqref{eq:Pb_2_MMSE}
can be approximated as $\psi\approx\left(N-K\right)\left(2+N/K\right)^{L-1}.$
Thus, 
\begin{equation}
\eta_{2}\approx\beta_{2}\lambda_{2}^{L},
\end{equation}
where $\beta_{2}=6\left(\eta p_{1}+m\right)\left(N-K\right)/\xi\Omega\left(2+N/K\right)$
which is a decreasing function of $L$ and $\lambda_{2}=\rho\left[1+K/\left(N+K\right)\right].$

For $M\ge4$, we have $\lambda_{2}\le\left[1+K/\left(N+K\right)\right]/2<1$
for any $K<N$. Hence, $\lim_{L\rightarrow\infty}\eta_{2}=\lim_{L\rightarrow\infty}\beta_{2}\lambda_{2}^{L}=0.$

For $M=2,$ we attain $\lambda_{2}=1+K/\left(N+K\right)>1$. Thus,
$\lim_{L\rightarrow\infty}\Delta_{2}=\left(10/L\right)\log_{10}\left[1+K/\left(N+K\right)\right]^{L}=10\log_{10}\left[1+K/\left(N+K\right)\right]$
(dB). Moreover, due to $1+K/\left(N+K\right)<1.5$, $\lim_{L\rightarrow\infty}\Delta_{2}<\left(10/L\right)\log_{10}\left(1.5^{L}\right)\approx1.76<\lim_{L\rightarrow\infty}\Delta_{1}\approx3$
(dB). 
\end{IEEEproof}
\textit{Remark 5.} Compared to the perfect CSI case (Theorem~1),
Theorem~3 indicates that for given $M$, the performance gap between
the two detectors under uncertain CSI gets smaller than that under perfect
CSI. Therefore, the GD becomes more attractive than the ML under the
MMSE CSI condition, particularly when the receiver has more antennas.

\section{Simulation Results}

We provide simulation results for MCIK-OFDM-SC having $N_{c}=128$ total
sub-carriers, which are divided into $G$ clusters, each having $N$
sub-channels. For illustrations, we consider $N\in\left\{ 2,4\right\} $,
$K<4$, $M\in\left\{ 2,4,8\right\} $, and $L\in\{1,2,4,8,12\}$.
The BER simulation results for the GD are compared to the ML under
various MCIK parameters and CSI conditions.

\subsection{Accuracy of Theoretical and Asymptotic Expressions}

\begin{figure}[tb]
\begin{centering}
\includegraphics[width=0.9\columnwidth]{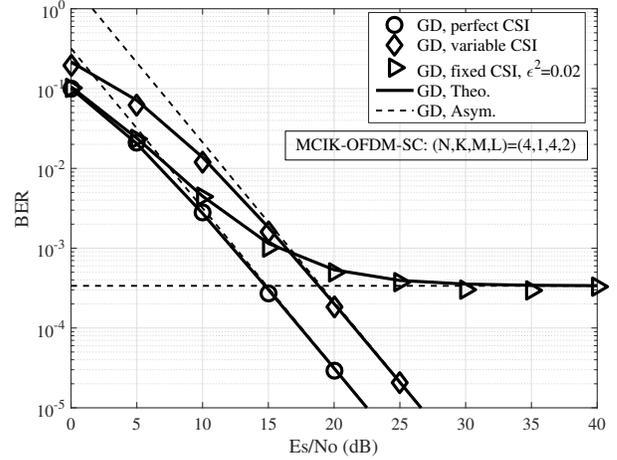}
\par\end{centering}
\caption{BER of the GD detector in MCIK-OFDM-SC under various CSI conditions, with $(N,K,M,L)=(4,1,4,2)$.
\label{fig:D01}}
\end{figure}
\begin{figure}[tb]
\begin{centering}
\includegraphics[width=0.9\columnwidth]{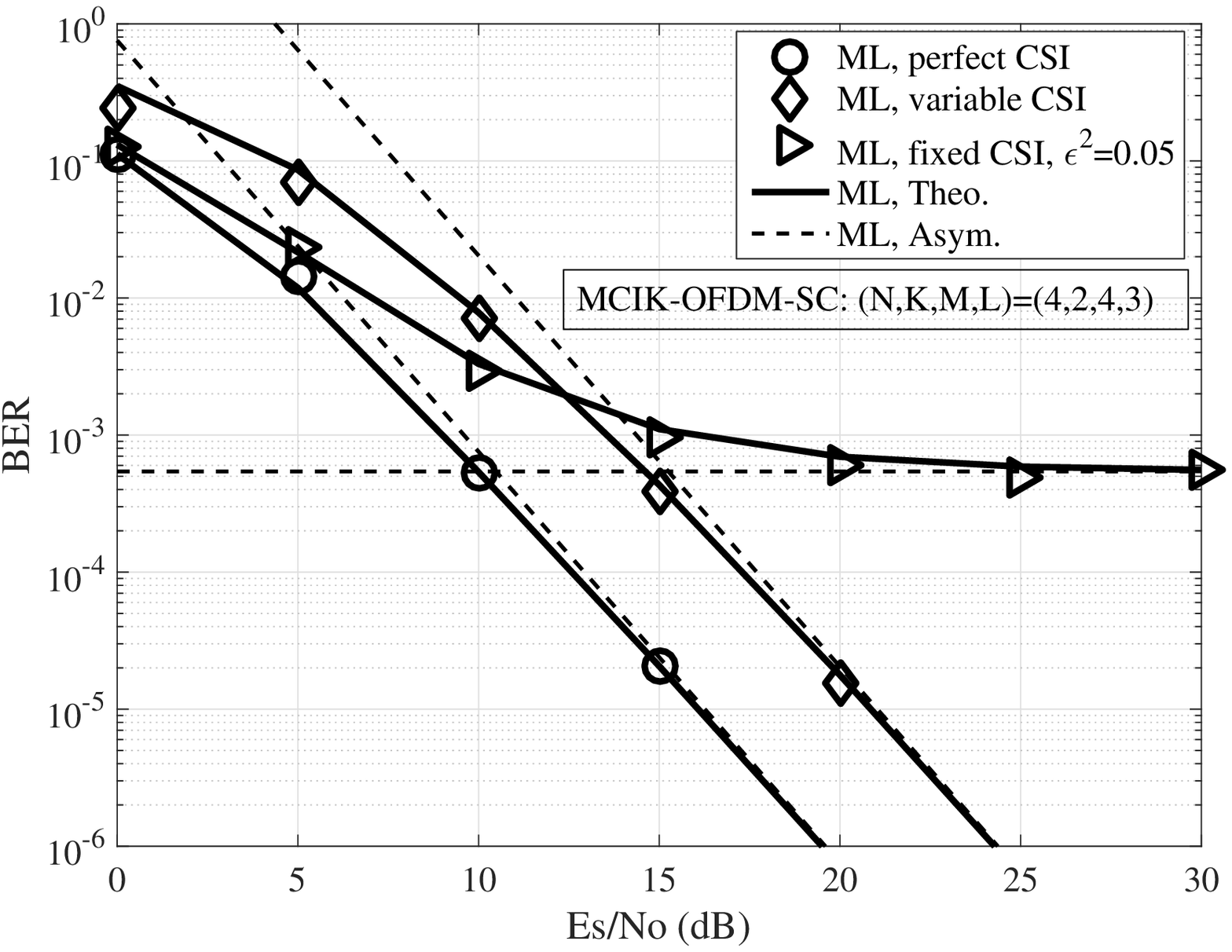}
\par\end{centering}
\caption{BER of the ML detector in MCIK-OFDM-SC under various CSI conditions, with $(N,K,M,L)=(4,2,4,3)$.
\label{fig:D03}}
\end{figure}
\begin{figure}[tb]
\begin{centering}
\includegraphics[width=0.9\columnwidth]{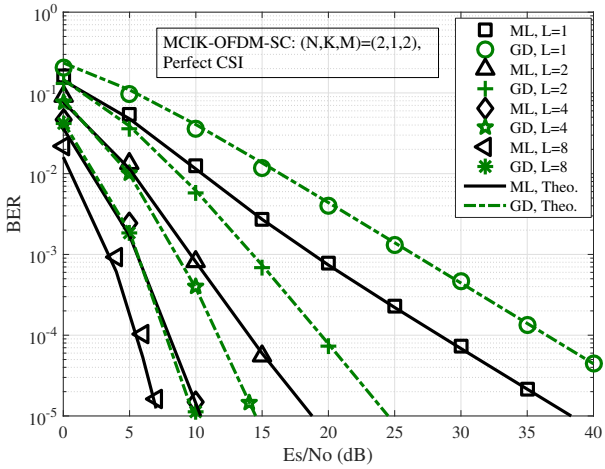}
\par\end{centering}
\caption{BER comparison between the ML and the GD in MCIK-OFDM-SC under perfect
CSI, with $(N,K,M)=(2,1,2)$ and $L=1,2,4,8$. \label{fig:A01}}
\end{figure}
Fig.~\ref{fig:D01} depicts the simulation results of MCIK-OFDM-SC
using the GD, along with the theoretical and asymptotic BER expressions
when $(N,K,M,L)=(4,1,4,2)$, under various CSI conditions. \textcolor{black}{As observed 
from Fig.~\ref{fig:D01}, the theoretical BER expressions derived for the GD 
   are very tight, i.e., very close to simulation results in a broad range of SNRs, while the asymptotic results
are accurate in high SNR regions. This observation clearly confirms the accuracy of our theoretical analysis provided in Section~III and Section~IV.} In addition, under fixed or variable
$\epsilon^{2}$, the GD suffers from a considerable loss in the BER
compared to the perfect CSI case ($\epsilon^{2}=0$). For example,
at BER of $10^{-3}$ in Fig.~\ref{fig:D01}, the loss of SNR gain
caused by fixed or variable CSI uncertainty is more than 4 dB. Note 
that a similar observation can be seen in Fig.~\ref{fig:D03} for
the ML detector.

\subsection{BER under Perfect CSI}

\begin{figure}[tb]
\begin{centering}
\includegraphics[width=0.9\columnwidth]{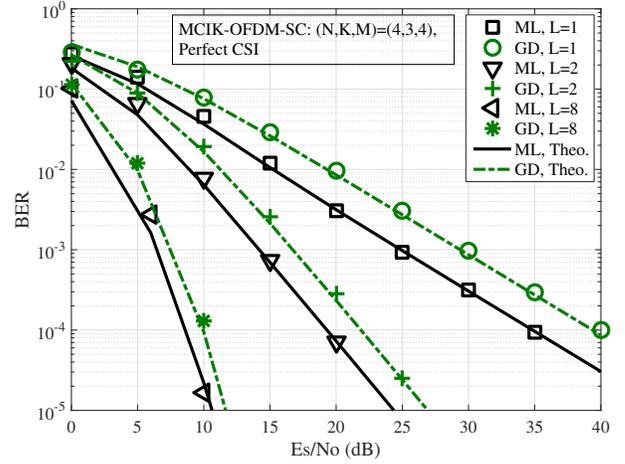}
\par\end{centering}
\caption{BER comparison between the ML and the GD in MCIK-OFDM-SC under perfect
CSI, with $(N,K,M)=(4,3,4)$ and $L=1,2,8$. \label{fig:A02}}
\end{figure}
Fig.~\ref{fig:A01} depicts the BERs for the ML and the GD in MCIK-OFDM-SC
under perfect CSI, with $(N,K,M)=(2,1,2)$ and $L=1,2,4,8$. As observed
from Fig.~\ref{fig:A01}, the ML always outperforms the GD even as
$L$ increases. For instance, as $L=8$, at BER of $10^{-4}$, the
ML achieves the SNR gain of 3 dB over the GD. This confirms Theorem~1
as $M=2$.

In Fig.~\ref{fig:A02}, the BER comparison between the two detectors
under perfect CSI is illustrated for MCIK-OFDM-SC with $(N,K,M)=(4,3,4)$
and $L=1,2,8.$ It is shown from Fig.~\ref{fig:A02} that when $M=4$,
the BER of the GD approaches to that of the ML as $L$ gets larger.
In particular, at BER of $10^{-3}$, the coding gain attained by the
ML over the GD is about 5 dB when $L=1$, while this gain reduces
to only 1 dB when $L=8$. This validates Theorem~1 for the case of
$M=4.$

\begin{figure}[tb]
\begin{centering}
\includegraphics[width=0.9\columnwidth]{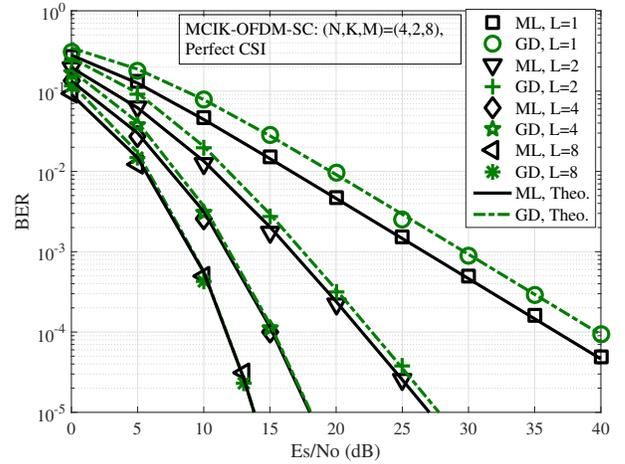} 
\par\end{centering}
\caption{BER comparison between the ML and the GD in MCIK-OFDM-SC under perfect
CSI, with $(N,K,M)=(4,2,8)$ and $L=1,2,4,8$. \label{fig:A03}}
\end{figure}

Fig.~\ref{fig:A03} illustrates the BERs for the ML and the GD when
$(N,K,M)=(4,2,8)$ and $L=1,2,4,8$. It is clear from this figure
that the BER of GD rapidly tends to that of the ML as $L$ increases.
Specifically, as $L=4$, the performance gap between these two detectors
becomes negligible. This confirms Theorem~1 for the case of $M\ge8.$

\subsection{BER under Fixed CSI Uncertainty}

\begin{figure}[tb]
\begin{centering}
\includegraphics[width=0.9\columnwidth]{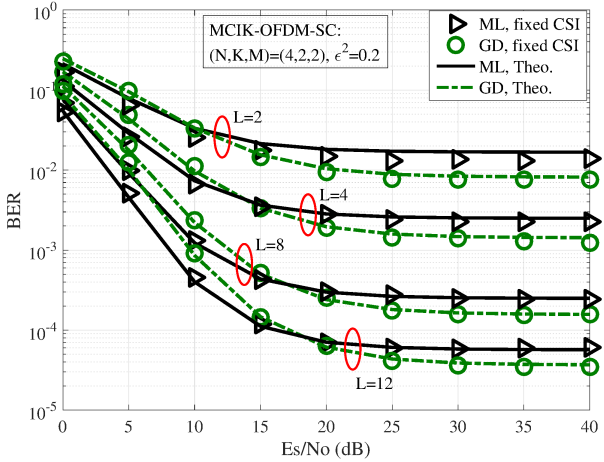} 
\par\end{centering}
\caption{BER comparison between the ML and the GD in MCIK-OFDM-SC under fixed
CSI, with $(N,K,M)=(4,2,2)$, $L=2,4,8,12$ and $\epsilon^{2}=0.2$.
\label{fig:E03}}
\end{figure}

Fig.~\ref{fig:E03} depicts the BER comparison between the ML and
the GD under fixed CSI uncertainty, with $(N,K,M)=(4,2,2)$, $L=2,4,8,12$
and $\epsilon^{2}=0.2$. \textcolor{black}{Interestingly, it can be seen from this figure that at high SNRs, the GD outperforms the ML in terms of the BER.
For example, as $L=4,$ the GD achieves the BER lower than the ML
when $E_{s}/N_{0}\ge15$ dB. This is due to the fact that under the fixed CSI uncertainty, using the energy detection, the GD achieves better index detection performance than its ML counterpart, leading to better BER performance, as theoretically proved in Subsection~IV-B.} Moreover, due to the fixed error variance,
i.e., $\epsilon^{2}=0.2$, there exists error floors on the BERs of
the two detectors. These floors get lower as $L$ increases. This
observations validate Theorem~2.

\subsection{BER under MMSE Variable CSI Uncertainty}

\begin{figure}[tb]
\begin{centering}
\includegraphics[width=0.9\columnwidth]{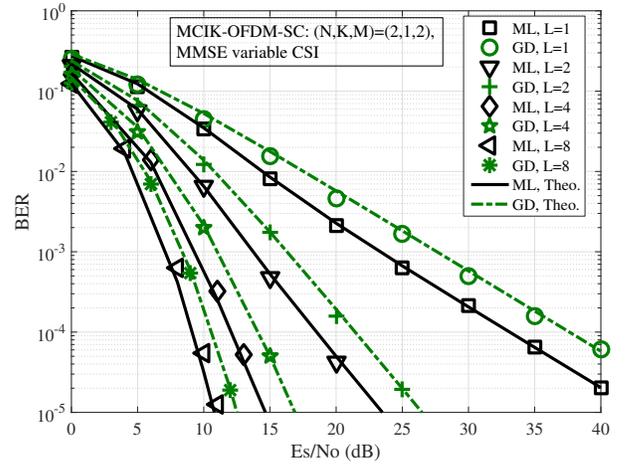} 
\par\end{centering}
\caption{BER comparison between the ML and the GD in MCIK-OFDM-SC under MMSE
variable CSI uncertainty, with $(N,K,M)=(2,1,2)$ and $L=1,2,4,8$.
\label{fig:C01}}
\end{figure}

Fig.~\ref{fig:C01} depicts the BER comparison between the GD and
ML detectors under MMSE-based variable CSI uncertainty, with $(N,K,M)=(2,1,2)$
and $L=1,2,4,8$. As seen via Fig.~\ref{fig:C01}, when $L$ gets
larger, the BERs of the two detectors become closer. However, the
ML always outperforms the GD. In addition, the performance gap between
them under variable CSI uncertainty gets smaller than that under perfect CSI in
Fig.~\ref{fig:A01}. These observations  validate Theorem~3 for $M=2$. 

\begin{figure}[tb]
\begin{centering}
\includegraphics[width=0.9\columnwidth]{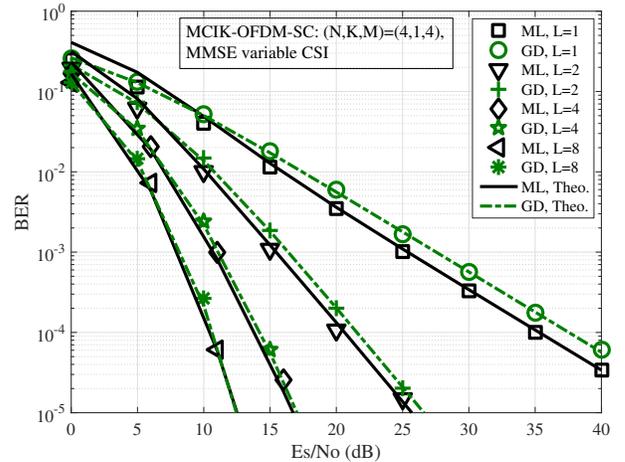} 
\par\end{centering}
\caption{BER comparison between the ML and the GD in MCIK-OFDM-SC under MMSE
variable CSI uncertainty, with $(N,K,M)=(4,1,4)$ and $L=1,2,4,8$.
\label{fig:C02}}
\end{figure}

Fig.~\ref{fig:C02} compares the BER between the two detectors
under MMSE variable CSI, when $(N,K,M)=(4,1,4)$ and $L=1,2,4,8$.
Unlike the perfect CSI case, the BERs of the ML and
the GD under this CSI condition quickly converge to each other as $L$ increases even when
$M=4$. Similar to Fig.~\ref{fig:A03}, as $L\ge2$ there is a marginal
gap in the BER between the two detectors. Hence, Theorem~3
with $M\ge4$ is clearly validated.

\section{Conclusions}

We proposed a generalized framework for the BER analysis of MCIK-OFDM
using either the GD or ML detector. Based on this, we derived tight, closed-form
expressions for the BERs of MCIK-OFDM with the selection combining
ML (or GD) receiver, taking effects of CSI uncertainty into account.
We provided the asymptotic analysis to investigate impacts of imperfect
CSI on their BERs. Furthermore, the BER comparison between
the GD and ML detectors under various CSI conditions was presented, which
allows to provide a theoretical guideline on the signal detection
of MCIK-OFDM-SC under each specific CSI condition. For example, under
MMSE-based variable CSI, the SC-based GD was shown to approach the
SC-based ML in terms of the BER as the number of antennas increases
and $M\ge4$. More interestingly, under fixed CSI uncertainty and
at high SNRs, the SC-based GD always outperforms the SC-based ML in
terms of the BER for any value of $M$. Finally, the derived BER expressions
and theoretical guideline are validated via simulation results. It
is noteworthy that the derived expressions and proposed guideline
for using the GD would be useful for various designs of the practical
implementation of MCIK-OFDM. \textcolor{black}{In our future work, we plan to investigate the performance of MCIK-OFDM-SC in combination with a number of diversity enhancement techniques, such as coordinate interleaving \cite{CIbasar2015}, repetition codes \cite{ThienTWC2018}, and spreading matrix \cite{ThienTVT2018}.}

\appendices{}

\section{Proof of Lemma 1}

The instantaneous SEP of the classical PSK symbol detection per sub-carrier
$\alpha$ (denoted by $P_{M}\left(\alpha\right)$) is given by \cite{thienBERGD}
\begin{equation}
P_{M}\left(\alpha\right)\approx\frac{\xi}{12}\left[e^{-\frac{\rho\bar{\gamma}\hat{\nu}_{\alpha}}{1+\epsilon^{2}\bar{\gamma}}}+3e^{-\frac{4\rho\bar{\gamma}\hat{\nu}_{\alpha}}{3\left(1+\epsilon^{2}\bar{\gamma}\right)}}\right],\label{eq:ins_PM}
\end{equation}
where $\xi=1$ for $M=2$ and $\xi=2$ for $M>2$, and $\hat{\nu}_{\alpha}=\left|\hat{h}\left(\alpha\right)\right|^{2}$
which is chi-square distributed with degrees of freedom of two, .i.e,
$\hat{\nu}_{\alpha}\sim\mathcal{X}_{2}^{2}.$ Note that $\left|\hat{h}\left(\alpha\right)\right|^{2}=\max_{l}\left|\hat{h}_{l}\left(\alpha\right)\right|^{2}$
and using the order statistics theory, the probability density function
(PDF) of $\hat{\nu}_{\alpha}$ is given as 
\begin{equation}
f_{\hat{\nu}}\left(x\right)=\frac{L}{a}e^{-\frac{x}{a}}\left(1-e^{-\frac{x}{a}}\right)^{L-1},\label{eq:PDF_v_hat}
\end{equation}
where $a=1-\epsilon^{2}$. Using \eqref{eq:PDF_v_hat}, the MGF of
$\hat{\nu}_{\alpha}$ can be obtained, after simple manipulations,
as 
\begin{equation}
\mathcal{M}_{\hat{\nu}}\left(t\right)=\frac{L!}{\prod_{l=1}^{L}\left(l-at\right)}.\label{eq:MGF_v_hat}
\end{equation}

Finally, applying the MGF approach to \eqref{eq:ins_PM} and using
\eqref{eq:MGF_v_hat}, the average SEP of \eqref{eq:ins_PM} is attained
as \eqref{eq:PM}.

\section{Proof of Lemma 2 }

Let $b=\left|\hat{h}\left(\alpha\right)\right|$ and $Z=\left|\left|\hat{h}\left(\alpha\right)\right|+\tilde{e}\left(\alpha\right)\right|^{2}$.
Assume that $\tilde{e}\left(\alpha\right)=c+jd$, where $c,d\sim\mathcal{N}\left(0,\epsilon^{2}/2\right)$,
we obtain

\begin{equation}
Z=(b+c)^{2}+d^{2}.
\end{equation}
Let $Z'=2Z/\epsilon^{2}=\left[\sqrt{2}(b+c)/\epsilon\right]^{2}+\left(\sqrt{2}d/\epsilon\right)^{2}.$
Due to $\sqrt{2}(b+c)/\epsilon\sim\mathcal{N}\left(\sqrt{2}b/\epsilon,1\right)$
and $\sqrt{2}d/\epsilon\sim\mathcal{N}\left(0,1\right)$, $Z'$ is
distributed according to the noncentral chi-squared distribution with
two degrees of freedom, i.e., $\mathcal{X}_{2}^{2}\left(\lambda\right)$,
where $\lambda=2b^{2}/\epsilon^{2}$ is the non-centrality parameter
\cite{johnson1995continuous}. Thus, the MGF of $Z'$ is given by
\cite{johnson1995continuous} 
\begin{equation}
\mathcal{M}_{Z'}\left(t\right)=\frac{e^{\frac{2b^{2}t/\epsilon^{2}}{1-2t}}}{1-2t}.\label{eq:MGF_Z_comma}
\end{equation}

Finally, the MGF of $Z$ can be computed, using $\mathcal{M}_{Z'}\left(t\right)$
in \eqref{eq:MGF_Z_comma} as $\mathcal{M}_{Z}\left(t\right)=\mathcal{M}_{Z'}\left(\epsilon^{2}t/2\right)$,
which leads to \eqref{eq:MGF_non_central}.

 \bibliographystyle{IEEEtran}
\phantomsection\addcontentsline{toc}{section}{\refname}\bibliography{Ref}

\end{document}